\documentclass[preprint,5p,twocolumn]{elsarticle}

\usepackage[ruled,linesnumbered]{algorithm2e} % For algorithms
\SetKwRepeat{Do}{do}{while}
\SetAlFnt{\small}
\usepackage{listings}
\usepackage{color}
\usepackage[table]{xcolor}
\usepackage{amsthm}
\usepackage{amsmath}
\usepackage{tabularx}
\usepackage{graphicx}
\usepackage{subcaption}
\usepackage{booktabs}
\usepackage[normalem]{ulem}
\usepackage{multirow}
\usepackage{rotating}
\usepackage{enumitem}
\usepackage[bookmarks=false]{hyperref}

\usepackage{framed}
\usepackage[many]{tcolorbox}
\newtcolorbox{framefloat}[1]{arc=0pt,outer arc=0pt,boxrule=0.5pt,
  colframe=black,colback=white,left=2mm,right=2mm}

\theoremstyle{definition}

\journal{Journal of Systems and Software}

\usepackage[T1]{fontenc}

\begin{document}

\begin{frontmatter}

\title{Software Runtime Monitoring with Adaptive Sampling Rate to Collect Representative Samples of Execution Traces}

\author[inst1]{Jhonny Mertz\corref{cor1}}
\ead{jmamertz@inf.ufrgs.br}
\cortext[cor1]{Corresponding author.}

\author[inst1]{Ingrid Nunes}
\ead{ingridnunes@inf.ufrgs.br}

\affiliation[inst1]{organization={Universidade Federal do Rio Grande do Sul (UFRGS)},%Department and Organization
            % addressline={Address One}, 
            city={Porto Alegre},
            % postcode={00000}, 
            % state={Rio Grande do Sul},
            country={Brazil}}

\begin{abstract}
Monitoring software systems at runtime is key for understanding workloads, debugging, and self-adaptation. It typically involves collecting and storing observable software data, which can be analyzed online or offline. Despite the usefulness of collecting system data, it may significantly impact the system execution by delaying response times and competing with system resources. The typical approach to cope with this is to filter portions of the system to be monitored and to sample data. Although these approaches are a step towards achieving a desired trade-off between the amount of collected information and the impact on the system performance, they focus on collecting data of a particular type or may capture a sample that does not correspond to the actual system behavior. In response, we propose an adaptive runtime monitoring process to dynamically adapt the sampling rate while monitoring software systems. It includes algorithms with statistical foundations to improve the representativeness of collected samples without compromising the system performance. Our evaluation targets five applications of a widely used benchmark. It shows that the error (RMSE) of the samples collected with our approach is 9--54\% lower than the main alternative strategy (sampling rate inversely proportional to the throughput), with 1--6\% higher performance impact.
\end{abstract}


\end{frontmatter}
% \linenumbers

% !TEX root = ../main.tex

\section{Introduction}\label{sec:introduction}

Software runtime monitoring~\cite{Gao2017} is fundamental for ensuring software quality~\cite{Kang2018}. It collects and often stores observable software data to enable, e.g., runtime verification, debugging, program comprehension, and self-adaptation~\cite{Feng2018}. A typical type of software runtime data consists of execution traces~\cite{Pirzadeh2011,Yuan2014,Reger2016}. An \emph{execution trace} is a record that provides information about the execution of software systems. These traces capture system operations, such as method calls or message exchanges, and include information of interest (e.g., method inputs and outputs) depending on the monitoring goal. Such goals can be the validation of quality requirements~\cite{Finocchi2013}, identification of security vulnerabilities~\cite{Yuan2014} or model inconsistencies~\cite{Bartocci2018}, performance engineering~\cite{DellaToffola2015,Mertz2018a}, and optimization~\cite{Feng2018}. AWS X-Ray\footnote{\url{https://aws.amazon.com/xray/}}, for example, is a tool that uses tracing execution to provide an end-to-end view of request paths in software applications, including a map of the application's underlying components. When an exception occurs while the application is serving an instrumented request, AWS X-Ray records details about the exception, including the stack trace. This helps, e.g., identify and troubleshoot the root cause of performance issues and errors.

Despite the usefulness of execution traces, collecting them at runtime consumes resources and may cause performance decays~\cite{Bartocci2018}, mainly when they include detailed information, such as method parameters. To address this, execution traces can be sampled or filtered. \emph{Filtering} and \emph{sampling} execution traces have been commonly adopted with pre-defined and fixed configurations, which specify certain software locations to be monitored and/or a sampling rate~\cite{Hamou-Lhadj2004,Pirzadeh2011,Pirzadeh2013,Las-Casas2018}. These configurations may be unsuitable to cope with software usage peaks and unable to handle unforeseen scenarios. AWS X-Ray, for instance, applies a conservative sampling strategy and records only the first request of each second and five percent of any additional requests. Any different strategy must be manually managed, considering the performance impact it may cause to the application. These limitations are addressed by \emph{adaptive} approaches~\cite{Zavala2019}. However, existing work either focuses only on collecting traces for a particular purpose~\cite{Fei2006,Las-Casas2018} or uses a strategy that cannot guarantee that the collected traces are a \emph{representative} sample of the population~\cite{Hauswirth2004,Bronink2016}. This can potentially cause wrong decision making based on the sample or missing information.

To address the challenge of obtaining a representative sample while monitoring software systems, we propose an adaptive sampling process to collect execution traces with detailed information in environments where the performance impact is critical, such as production environments. Our goal is to pursue the representativeness of the samples of execution traces while adjusting the sampling rate used to monitor a software system to cope with increases in its workload. The process decides whether the operations executed to respond to each incoming request should be recorded as execution traces. Our process is composed of three activities: (1) \emph{sampling decision}, which decides whether a request (with associated execution traces) should be recorded and included in a sample; (2) \emph{sampling rate adaptation}, which adjusts the sampling rate at runtime; and (3) \emph{sample evaluation}, which assesses the representativeness of the sample to identify the end of the monitoring cycle. These activities include algorithms with statistical foundations to ensure that, at the end of each monitoring cycle, the collected sample is representative of the population. Our evaluation targets five applications of the well-known DaCapo benchmark, which includes workloads that are used to create usage variations that occur in production environments such as stationary usage and spikes. The root-mean-square error (RMSE) of the samples collected with our approach is 9--54\% lower than the main alternative strategy (sampling rate inversely proportional to the throughput), with 1--6\% higher performance impact.

Because the intervention made in our evaluation consists solely in varying sampling rates, our conclusion provides reliable evidence that our adaptive sampling process is effective and efficient to be used while monitoring software systems. Our proposed process can thus be integrated as part of monitoring solutions, which can also include other mechanisms, such as filtering. The choice for suitable mechanisms for monitoring a particular software system depends, however, on the monitoring goal, e.g.\ runtime adaptation or anomaly detection.

The remainder of the paper is organized as follows. Next, we discuss existing software monitoring approaches. Our proposed solution is introduced in Section~\ref{sec:proposed_solution} and its empirical evaluation is presented in Section~\ref{sec:evaluation}. Finally, we conclude in Section~\ref{sec:conclusion}.

% !TEX root = ../main.tex

\section{Related Work}
\label{sec:related_work}

Monitoring executions traces with detailed information requires code instrumentation~\cite{Cassar2017}, has a performance impact, and often consumes extensive storage space~\cite{Mertz2017b}. Filtering and sampling have been demonstrated as practical mechanisms to reduce the monitoring overhead and enable faster trace analysis~\cite{Bartocci2018}. While \emph{filtering} excludes from the monitoring particular executions or software locations that are not of interest for a given goal (e.g.\ troubleshoot performance issues or find bugs), \emph{sampling} involves establishing a sampling rate to monitor a subset of execution traces assuming that it is a representative sample of the population of traces. These two approaches can be combined to keep the monitoring overhead at acceptable levels. There is existing work to support the specification of filtering and sampling configurations, which give the scope of monitoring (i.e.\ locations to monitor) and the sampling rate, respectively. Proposed approaches can be classified into two groups: (i) \emph{fixed configuration}, which keeps the same configuration throughout the software execution until it is manually updated; (ii) \emph{adaptive configuration}, which dynamically adjusts the configuration based on constraints and lightweight monitored data. These groups are discussed as follows.

\subsection{Fixed Configuration}

A straightforward way to cope with the monitoring overhead is to use random or systematic sampling~\cite{Chan2003,Dugerdil2007,Jung2014,Zhou2016,Song2017}, which is used in various commercial and open-source tools~\cite{VanHoorn2012,Horky2016}. However, as there are traces that are not recorded, important traces may be missed. Thus, choosing a sampling rate is a challenge~\cite{Las-Casas2018,Miranskyy2016,Mertz2019,Mertz2020} because of the trade-off between the representativeness of the sample and the performance overhead. A suitable solution in some scenarios is the use of a fixed (higher) sampling rate but targeting particular executions or regions of an application. Nevertheless, when the population of traces is not homogeneous, focusing on statically defined regions may lead to reduced coverage and thus an unrepresentative sample~\cite{Pirzadeh2013}.

To support specifying a filtering configuration, there are approaches that perform an automated offline analysis of the program to define relevant application regions or paths~\cite{Apiwattanapong2003,Santelices2006,Sridharan2007,Narayanappa2010}. Although helpful, these solutions are not suitable when the areas of interest vary at runtime. In these cases, the approaches should re-executed to tune the monitoring configuration.  

\subsection{Adaptive Configuration}

Adaptive monitoring approaches change, at runtime, the sampling and filtering configuration and even collected metrics. Work in this direction has been recently investigated in a systematic mapping~\cite{Zavala2019} that reveals that most of the proposed adaptations focus on improving the monitoring results for a \emph{specific purpose}. Fei~\cite{Fei2006}, for example, observed that executing a (region of an) application with the same context tends to produce the same outcome. Therefore, repeated executions do not need to be monitored when the goal is to identify where bugs are likely to occur. Similarly, Las-Casas~\cite{Las-Casas2018,Las-Casas2019} aim at capturing outliers and anomalous traces. They aim to maximize the diversity of execution traces in the sample with infrequent patterns by computing the distance among traces and ensuring diversity in the set of traces, given a fixed budget. Targeting performance, Ding~\cite{Ding2015} proposed a cost-aware logging mechanism that decides whether to keep log messages based on (1) a dynamic measurement of the performance of the code snippet that generated the log and (2) an allowed maximum volume of logs in a time interval. The goal is to keep logs related to code snippets that execute slower than in the past. In a different direction, \emph{Tigris}~\cite{Mertz2019,Mertz2020}, a two-phase monitoring framework, can be instantiated considering a specific monitoring goal. The framework includes a domain-specific language to specify the criteria used in the first phase of the framework to dynamically choose the application locations that should be monitored in detail in the second phase.

These aforementioned adaptive approaches focus on dynamic filtering and adopt a fixed sampling rate---some~\cite{Fei2006, Mertz2020} allow it to be given as a parameter. There are three approaches that adjust the sampling of the execution traces to cope with the monitoring performance overhead. Focusing on memory, Daoud~\cite{Daoud2017} proposed a dynamic tracing approach that monitors memory usage to decide whether to collect a trace based on pre-defined conditions and thresholds such as the time elapsed since the last trace or the amount of memory allocation calls. Targeting the application workload, two approaches~\cite{Hauswirth2004,Bronink2016} propose to use a sampling rate that is inversely proportional to the frequency of execution, which gives the application throughput. Although this strategy is able to reduce the monitoring overhead when the application is overloaded, the collected sample may not correspond to the population of execution traces. In usage peaks, the proportion of collected execution traces is smaller than in typical workloads. As result, collected samples are likely not representative considering the population.

In summary, monitoring with a fixed configuration is suitable for software applications that have a workload with low variance, but may lead to an unacceptable overhead when there are peaks and cannot deal with unpredicted situations. Existing approaches dynamically adapt monitoring configurations by filtering traces or adjusting how the sampling is performed. The former assumes that there is a subset traces that are of interest and aim at collecting only them. The latter aims at having spacial coverage (that is, to monitor the whole application) but compromise temporal coverage by dynamically changing the sampling rate to keep the monitoring at an acceptable level. Our goal, similarly to Hauswirth~\cite{Hauswirth2004} and Bronink~\cite{Bronink2016}, is to collect a sample of execution traces with spacial coverage. However, we propose a decision-making process to collect execution trace samples that are \emph{representative} of the population. Our process is described in the next section, followed by an evaluation that compares our approach to their strategy (sampling rate inversely proportional to the throughput) and uniform sampling.

% !TEX root = ../main.tex

\section{Adaptive Sampling Process}
\label{sec:proposed_solution}

Given the limitations of existing work and our goal of collecting representative samples of detailed execution traces at runtime with controlled performance impact, we propose a three-activity process for monitoring software applications with an adaptive sampling rate. Our decisions are at the granularity of application request, which have method calls (executed to respond to it) recorded as detailed execution traces. This occurs if the request is selected to be part of a sample collected in a monitoring cycle. We use the \emph{PetClinic}\footnote{\url{https://projects.spring.io/spring-petclinic/}} project as a running example to explain the activities of our process. It is a web application that demonstrates the use of the Spring Framework and its features. It provides features (possible requests) in which employees of a pet clinic can view and manage information regarding veterinarians (\texttt{/vets}), clients (\texttt{/owners}), and their pets (\texttt{/pets}). It also includes a home page (\texttt{/home}), which is the entry point for users. We next first overview our process and its activities, and then describe each activity in detail.

\subsection{Process Overview}

\begin{figure}[t]
\centering
\includegraphics[width=\linewidth]{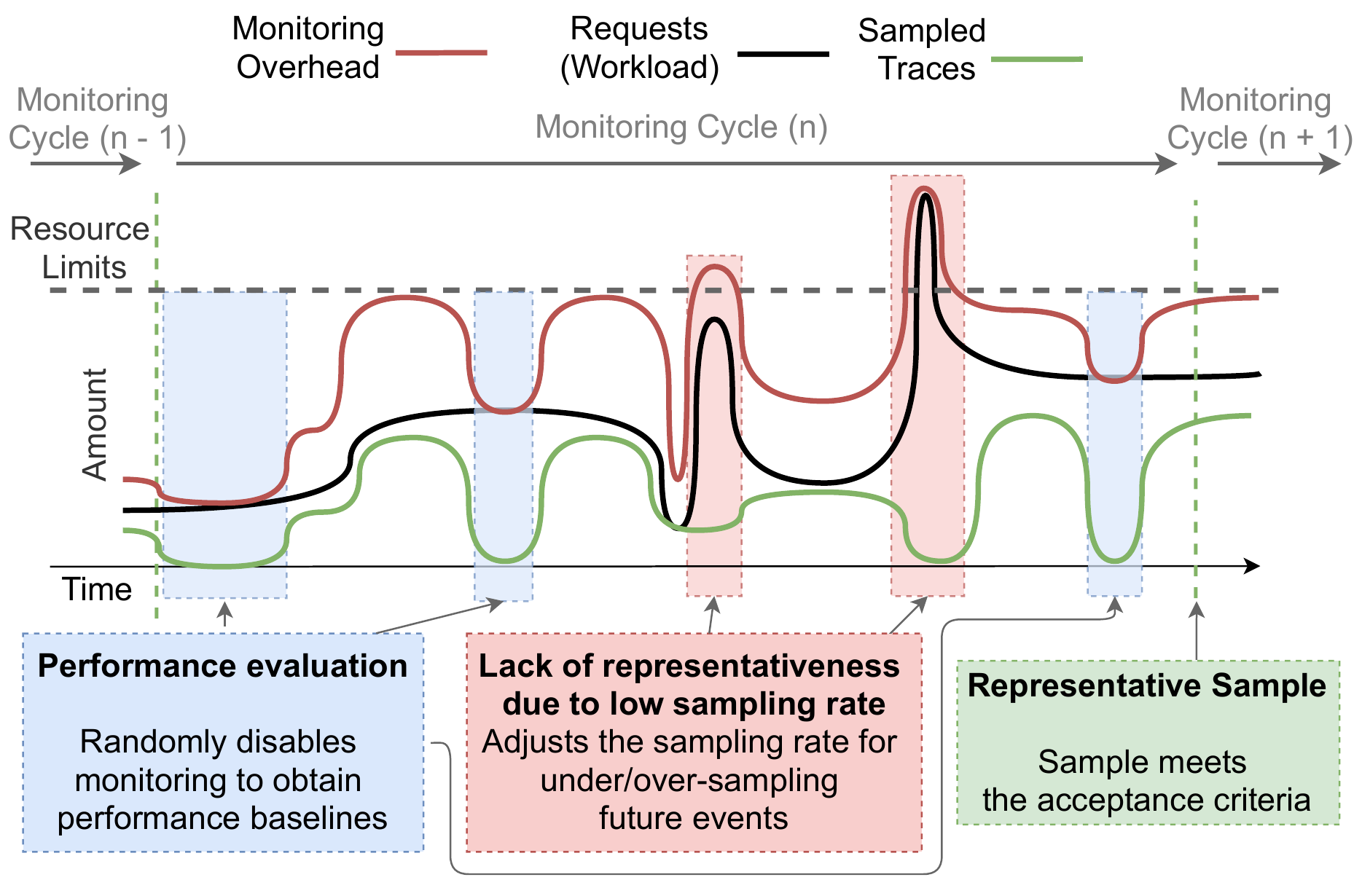}
\caption{Illustration of the Adaptive Sampling Process in Action: The figure shows how the sampling rate varies (in terms of the number of sampled traces) according to the current application workload. In peaks, the sampling rate is decreased to reduce the monitoring overhead.}
\label{fig:solution_timeline}
\end{figure}

The key idea underlying our process is to decrease the monitoring overhead to an acceptable level when the target software application needs its resources for regular processing and increasing it after the situation has been normalized. At the same time, we keep track of general statistics about the sample and population of requests to identify when a sample is representative. Our monitoring process is performed in \emph{cycles} and the result of each cycle is a representative sample. This behavior is shown in Figure~\ref{fig:solution_timeline}, where the black line represents the application workload, the red line represents the monitoring overhead, and the green line indicates the amount of requests and their execution traces being collected over time. 

\begin{figure}[t]
\centering
\includegraphics[width=\linewidth]{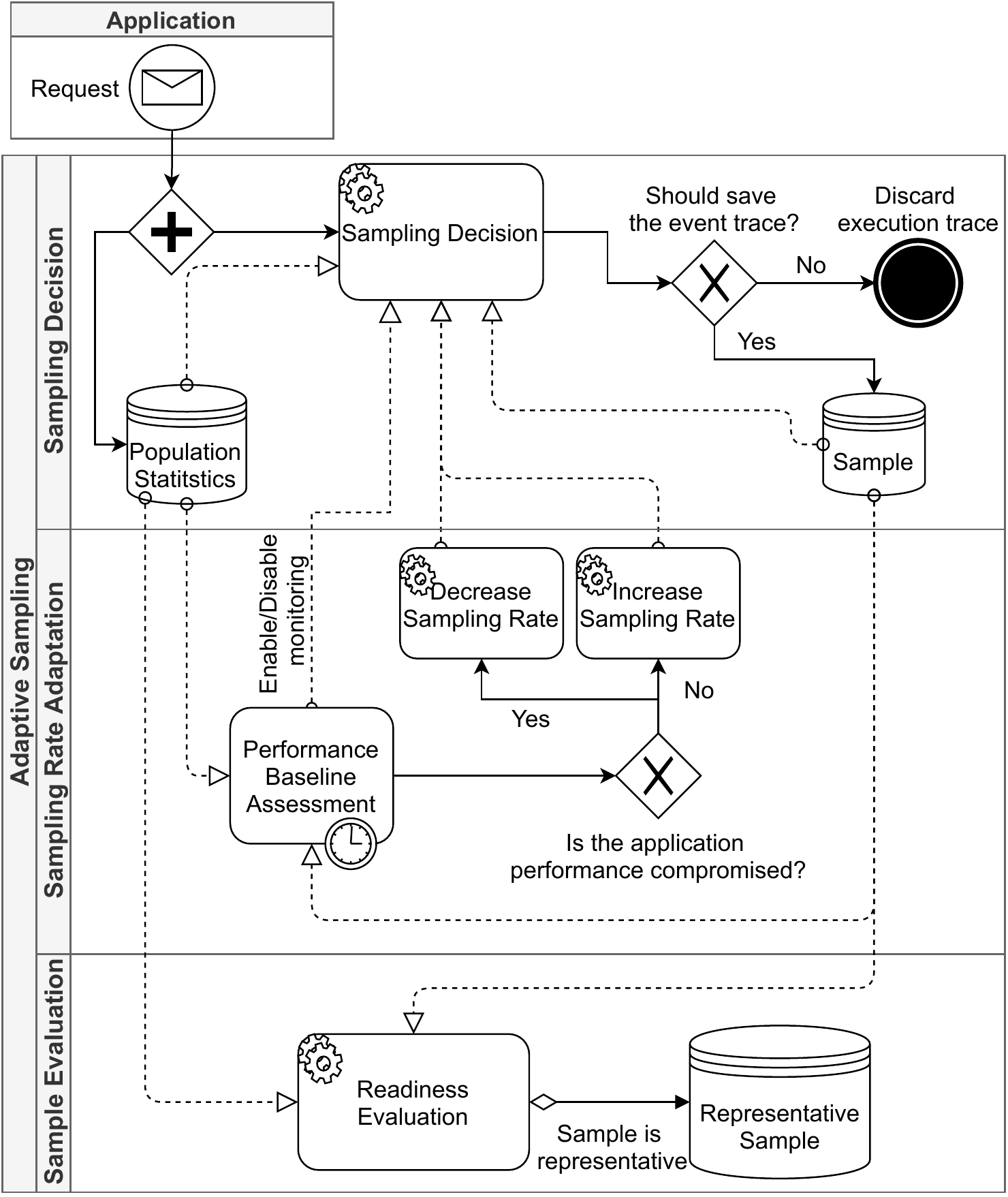}
\caption{Overview of our Adaptive Sampling Process.}
\label{fig:solution_overview}
\end{figure}

In order to make this behavior possible, three activities are performed in parallel at runtime as part of our process, as shown in Figure~\ref{fig:solution_overview}. The first activity, \emph{Sampling Decision}, is responsible for deciding whether an application request should have its associated execution traces collected and stored in detail (which is costly and requires I/O), taking into account both the sampling rate and the representativeness of the sample compared to the population. The sample representativeness is based on the distribution of application requests. All requests go through screening and by doing so, we also keep general statistics of the population of requests.

The sampling rate used in the Sampling Decision activity is updated by the \emph{Sampling Rate Adaptation} activity. It observes the current application workload as well as the monitoring overhead to decrease or increase the sampling rate. These adjustments of the sampling rate can be seen in Figure~\ref{fig:solution_timeline}---the larger the space between the black and red lines, the higher the sampling rate and, consequently, the monitoring overhead. When a performance degradation is perceived, i.e.\ the application response time increases, the sampling rate is decreased proportionally to the perceived degradation. To exemplify, suppose that an unexpected increase in the application workload results in the application bumping in the limits of its runtime platform (shadowed red boxes). In this situation, the monitoring is gradually reduced to allow the application to maintain its throughput. The sampling rate can be restored when the application workload decreases. To be able to assess the impact caused by monitoring, this activity also involves the collection of a performance baseline, which is a measurement of the application performance without monitoring. In Figure~\ref{fig:solution_timeline}, this occurs in the shadowed blue boxes.

Finally, the \emph{Sample Evaluation} activity is responsible for continuously evaluating the sample being collected to identify when it is considered representative. When the sample satisfies acceptance criteria with respect to the representativeness of the population, it is released for analysis, and a new monitoring cycle starts. If our monitoring process is used, e.g., in a self-adaptive system, the analysis of the sample can take place right after each cycle is concluded. After broadly understanding how our process works, we next describe each of its activities.

\subsection{Activity 1: Sampling Decision}
\label{subsection:sampling-decision}

The \emph{Sampling Decision} activity involves the execution of Algorithm~\ref{algorithm:sampling-decision} whenever a new observable request happens in the application. It performs three main tasks: (i) store statistics of the population of application requests (line 1); (ii) decide whether a request should be recorded with execution traces (lines 2--13) and (iii) store statistics of the request being added to the current sample (line 8), when applicable.

\begin{algorithm}[t]
\SetAlgoLined
\KwIn{$request$ to be processed by the application;}
\KwIn{current sampling rate $rate \in (0,1)$;}
\KwIn{monitoring control $isMonitoringEnabled$;}
\KwData{$population$, $sample$;}
\KwResult{$True$ if the execution traces of $request$ should be recorded, $false$ otherwise.}
\BlankLine
\BlankLine
add($request$, $population$)\;
\If{$isMonitoringEnabled$}{
    $shouldSample \leftarrow$ {\rm Bernoulli}($rate$)\;
    \If{$shouldSample$}{
        $E_{p}\leftarrow\{\, x \in population \mid x = request.id \,\}$\;
        $E_{s}\leftarrow\{\, y \in sample \mid y = request.id \,\}$\;
        \BlankLine
        \If{$\left(\dfrac{|E_{p}|}{|population|} \geq \dfrac{|E_{s}|}{|sample|}, \epsilon\right)$}{
                \BlankLine
                add($request$, $sample$)\;
                return $true$\;
        }
    }
}
return $false$\;
\caption{Sampling Decision}
\label{algorithm:sampling-decision}
\end{algorithm}

Storing statistics associated with the population and sample means keeping track of the \emph{frequency distribution} of each request in these two sets. This information is used to decide whether a particular request should be recorded with execution traces or whether a sample is representative and the monitoring cycle is complete. By adding requests to the population and sample (lines 1 and 8), we keep the number of each possible request registered, as shown in Table~\ref{tab:frequency_distribution} considering the PetClinic example. These values are obtained with an initial sampling rate of 50\%.

\begin{table}
\centering
\caption{Running example: Frequency distribution of the population and sample in the sampling decision activity.\label{tab:frequency_distribution}}
\begin{tabular}{lrrrr}
    \toprule
    \textbf{Request} & \multicolumn{2}{c}{\textbf{\textit{population}}} & \multicolumn{2}{c}{\textbf{\textit{sample}}} \\ 
    \midrule
    \texttt{/home}   & 105 & (47.7\%) & 53 & (47.7\%) \\
    \texttt{/vets}   & 43 &  (19.5\%) & 22 & (19.8\%)  \\
    \texttt{/pets}   & 62 &  (28.3\%) & 31 & (27.9\%) \\
    \texttt{/owners} & 10 &  (4.5\%)  & 5 & (4.5\%) \\ 
    \midrule
    \textbf{Total} & \textbf{220} & \textbf{(100\%)} & \textbf{111} & \textbf{(55.5\%)} \\ 
    \bottomrule
    \end{tabular}
\end{table}

A request is added to the sample and recorded with execution traces when three conditions are satisfied. The first refers to whether the monitoring is enabled (line 2). As introduced in the previous section, there are moments when the monitoring is disabled to obtain a performance baseline---this is controlled by the Sampling Rate Adaptation activity. The second condition involves randomly deciding whether the request should be selected based on the current sampling rate (\textit{rate}), which gives the probability of selecting a request as part of the sample. This decision is made using the Bernoulli distribution with parameter $p \in (0,1)$ to assign the value 1 (\textit{true}) with probability \textit{p} and the value 0 (\textit{false}) with probability $1 - p$ to the \textit{shouldSample} variable (line 3). The request satisfies the second criteria when \textit{shouldSample} is \textit{true}.

The third condition is related to the representativeness of the sample---we aim to keep its frequency distribution similar to that of the population (lines 5--7). The rationale is to not miss less frequent requests with, e.g., anomalies and exceptions. The idea is related to stratified sampling, where a population can be partitioned into subpopulations, i.e.\ clusters~\cite{Pirzadeh2011}, and a representative sample has the same class distribution as the population. The verification that is performed consists of a runtime \emph{resampling strategy} to balance the sample's class distribution according to the population's class distribution. This strategy is inspired by data analysis resampling to deal with unbalanced datasets~\cite{Estabrooks2004}. By resampling, we balance the sample classes according to the population distribution to keep representativeness in terms of proportion. It consists of ignoring new requests from the majority classes to allow minority classes to include more requests and increase their cardinality. As previously said, we keep the statistics of the population (\textit{population}) and sample (\textit{sample}) (exemplified in Table~\ref{tab:frequency_distribution}). These are kept as key-value mappings, where the type of a request is the key (\textit{request.id}), and its execution frequency within a monitoring cycle is the value. Using these statistics, it is possible to compute $E_{p}$ and $E_{s}$, which are the amount of requests of a particular type in the \textit{population} (line 5) and in the current \textit{sample} (line 6), respectively. Based on these values, we test if the sample is lacking requests of the type in consideration, considering an error margin $\epsilon$. For example, if a request of type \texttt{/vets} occurred (Table~\ref{tab:frequency_distribution}), it will be not added to the sample because it already has enough traces from \texttt{/vets} (19.8\%) when compared to the population (19.5\%). When a request satisfies this condition (line 7), the request is added to the sample (line 8) and the algorithm returns \textit{true} (line 9), that is, the request should be recorded with execution traces. If any of the three conditions is not satisfied, the algorithm returns \textit{false} and the request is not recorded, implying no additional monitoring overhead.

\subsection{Activity 2: Sampling Rate Adaptation}
\label{subsection:sampling-rate-adaptation}

The sampling rate used in the previously described activity is updated by Algorithm~\ref{algorithm:sampling-adaptation} executed in the Sampling Rate Adaptation activity. This is done considering the following premisses.

\begin{algorithm}[t]
\SetAlgoLined
\KwIn{current $sample$ being collected in the monitoring cycle;}
\KwIn{the maximum time window in which a performance baseline must be kept $\mathit{duration}$;}
\KwIn{the current performance $\mathit{currentPerf}$;}
\KwIn{monitoring control $\mathit{isMonitoringEnabled}$;}
\KwData{the current $\mathit{samplingRate}$; $\mathit{performanceRef}$;}
\KwResult{The updated sampling rate.}
\BlankLine
\BlankLine

addPerformanceSample($\mathit{performanceRef}$, $\mathit{currentPerf}$, $\mathit{isMonitoringEnabled}$)\;
$\mathit{medianRps} \leftarrow {\rm median}( \{ x \in \mathit{performanceRef}[RpS] \mid \mathit{performanceRef}[ME = isMonitoringEnabled] \})$\; 
$\mathit{normalBehavior} \leftarrow \mathit{performanceRef}[RpS = medianRpS \wedge ME = \mathit{isMonitoringEnabled}]$\;
$\mathit{equal} \leftarrow {\rm ttest}(\mathit{normalBehavior}, \mathit{currentPerf}, 0.05)$\;
$\mathit{diff} \leftarrow \dfrac{\sum_{i \in \mathit{currentPerf}} i}{\sum_{i \in normalBehavior} i} - 1$\;
\eIf{$\mathit{isMonitoringEnabled}$}{
    \eIf{equal or diff > 0}{
        $rate \leftarrow {\rm min} ( rate + (rate * \left|\mathit{diff}\right|) , maxRate )$\;
    }{
        ${\rm enablePerformanceBaseline}(duration)$\;
    }
}{
    \If{not equal and diff < 0}{
        $rate \leftarrow {\rm max} ( rate - (rate * \left|\mathit{diff}\right|) , minRate )$\;
    }
}
return $rate$\;
\caption{Sampling Rate Adaptation}
\label{algorithm:sampling-adaptation}
\end{algorithm}

\begin{enumerate}
	\item A software engineer is able to provide a desired sampling rate that leads to an \emph{acceptable performance impact} caused by monitoring when the application \emph{workload is typical}.
	\item The acceptable performance impact is not in terms of percentage but the absolute increase in the response time. For example, if the response time of a request is typically 100ms and with monitoring 105ms, the acceptable performance impact is 5ms and not 5\% of overhead.
	\item The sampling rate should be reduced to prevent an increase in the response time when the application is under stress, limited by a minimum required sampling rate.
	\item The sampling rate should be increased if it is below the desired level and the software application is returning to its typical behavior after a peak.
\end{enumerate}

Based on these premisses, the adaptation of the sampling rate requires three inputs: (i) \textit{maxRate}, which is the desired sampling rate and a higher sampling rate is not needed; (ii) \textit{minRate}, which is the minimum required sampling rate; (iii) the \textit{duration} of the period in which the approach should collect data to understand the application performance with the monitoring disabled (this is required for assessing the monitoring performance impact); and (iv) the \textit{frequency} in which the sampling rate is revised (in seconds).

Algorithm~\ref{algorithm:sampling-adaptation} performs the following tasks. First, it stores statistics of the application performance (line 1).
% INGRID: I wonder if a particular request was not called since the last execution
% JHONNY: We do not consider this request in the comparison then. Only those who appear on both sets
% INGRID: Ok, if a reviewer asks this, we clarify.
% INGRID: por que operations por segundo e não requests por segundo? Operations seria o quê? Method calls?
% JHONNY: operations = requests, I changed all ops -> rps and operations -> requests
The parameter \textit{currentPerf} gives the current application performance as a record with the number of requests executed since the last algorithm execution (RpS) and the average response time of each executed request. These data are stored in the \textit{performanceRef} table, which consists of a performance sample. For each record, we also store a flag indicating if these data correspond to a period in which monitoring is enabled. An example of \textit{performanceRef} is shown in Table~\ref{tab:performance_collection_example} for PetClinic. To avoid bias towards past observations and consuming unnecessary resources, \textit{performanceRef} is size-limited and stores the most recent executions, i.e.\ the oldest record is discarded to store a new one when the size limit is reached.

\begin{table}
\centering
\caption{Running example: \textit{performanceRef} table containing the response time of each request according to a given workload in requests per second (RpS). The ME column indicates whether the record was collected when monitoring was enabled. Rows highlighted in gray refer to the median of each group (ME = true and ME = false).\label{tab:performance_collection_example}}
\setlength{\tabcolsep}{1.1mm}
\begin{tabular}{lrrrrrl}
    \toprule
    \textbf{\#} & \textbf{RpS} & \textbf{/home} & \textbf{/vets} & \textbf{/pets} & \textbf{/owners} & \textbf{ME} \\ \midrule
    1 & 500 & 325ms & 450ms & 800ms & 1200ms & true \\
    2\cellcolor{gray!25} & \cellcolor{gray!25} 1500 & \cellcolor{gray!25} 400ms & \cellcolor{gray!25} 550ms & \cellcolor{gray!25} 900ms & \cellcolor{gray!25} 1500ms & true \cellcolor{gray!25} \\
    3\cellcolor{gray!25} & 2500\cellcolor{gray!25} & 600ms\cellcolor{gray!25} & 780ms\cellcolor{gray!25} & 1050ms\cellcolor{gray!25} & 1100ms\cellcolor{gray!25} & false \cellcolor{gray!25} \\
    4 & 550 & 350ms & 400ms & 650ms & 900ms & true \\
    ... & ... & ... & ... & ... & ... & ... \\
    $n-1$ & 325 & 430ms & 420ms & 480ms & 700ms & false \\
    $n$ & 200 & 270ms & 200ms & 235ms & 500ms & true \\ \bottomrule
    \end{tabular}
\end{table}

Based on these statistics, it is possible to derive both the typical application workload \textit{medianRpS} (line 2) and the corresponding normal behavior \textit{normalBehavior} in terms of response time (line 3), with and without monitoring. The typical application workload is given by the median of requests per second \textit{medianRpS} of the records in \textit{performanceRef}. Considering our example, these are the records \#2 and \#3, with and without monitoring, respectively. Because we need to select a single record to be used in the next algorithm tasks, if \textit{performanceRef} contains an even number of records, we select the highest requests per second (as opposed to the arithmetic mean of the two middle values) and the associated record, which is the \textit{normalBehavior}.

Then, we test if the current application performance \textit{currentPerf} is significantly different from the normal behavior \textit{normalBehavior} of the application to detect performance variations (line 4). This is done by comparing the averages of the execution times of each request type with a paired t-test, with the null hypothesis that the mean of the paired differences between \textit{currentPerf} and \textit{normalBehavior} is 0, with a significance level of 95\% ($p = 0.05$). Assume that Algorithm~\ref{algorithm:sampling-adaptation} is executing in our example with monitoring disabled and thus the \textit{normalBehavior} is record \#3. Let \textit{currentPerf} be
\begin{equation}
\{ \langle /home, 500 \rangle , \langle /vets, 720 \rangle , \langle /pets, 950 \rangle , \langle /owners, 1020 \rangle \}.
\end{equation}
The result of the comparison of \textit{normalBehavior} and \textit{currentPerf} is assigned to the variable \textit{equal}, which is the result of
\begin{equation}
ttest(\langle 600, 780, 1050, 1100 \rangle , \langle 500, 720, 950, 1020 \rangle , 0.05).
\end{equation}
This results in $equal = true$, indicating that there is no significant difference between these two groups. This is the first indicator of whether the sampling rate should be updated. The second indicator is the difference \textit{diff} (in percentage) between \textit{currentPerf} and \textit{normalBehavior} (line 5). In the example, it is 
\begin{equation}
\mathit{diff} = (3190 / 3530) - 1 = 0.9037 - 1 = -0.0963
\end{equation}
This means that the current performance is 9.63\% lower than the normal behavior.

These two indicators (\textit{equal} and \textit{diff}) are used together with the monitoring state of the application to decide whether the sampling rate should be updated. The sampling rate is increased---proportionally to the observed difference, limited by \textit{maxRate}---if the current performance (which includes monitoring) is similar or better than the normal behavior (lines 7--8). This means that the current execution times of the requests are similar or faster than the past observations, and thus the monitoring overhead is acceptable. The sampling rate is decreased (also based on \textit{diff}), if the normal behavior is significantly different from the current performance (which does not include monitoring, limited by \textit{minRate}) and the current performance is worse than the normal behavior (lines 13--15). This case occurs when the current execution times of the requests are slower when compared to past observations. In this situation, the application is (a) being impacted by users with an increased workload, or (b) the monitoring overhead is impacting the performance above acceptable levels. Note that $minRate > 0$ because, if the sampling rate reaches 0, it remains 0 indefinitely. 

In order to identify whether the application performance still faces degradation regardless of the case, there is a need for collecting a performance baseline (line 10). This occurs when the application is being monitored and the current performance is worse (significantly different and lower) than the normal behavior. In this case, the monitoring is then globally disabled and the performance baseline is collected according to the specified \textit{duration}. Finally, if none of these conditions are met, the sampling rate remains the same.

\subsection{Activity 3: Sample Evaluation}
\label{subsection:sample-readiness}

Our process aims at collecting a representative sample in each monitoring cycle. Therefore, when a new request is added to the sample, the sample representativeness is evaluated to determine if it is ready for being used. This is done in the Sample Evaluation activity that involves the execution of Algorithm~\ref{algorithm:sampling-evaluation}. A sample is considered representative if it satisfies three criteria: (i) it is larger than the minimum sample size; (ii) the performance between the sample and population is equivalent; and (iii) the sample distribution is similar to the population distribution. When the sample satisfies these criteria, there is statistical evidence that it is representative.

\begin{algorithm}[t]
\SetAlgoLined
\KwIn{the length of the current monitoring cycle $t$  in seconds;}
\KwData{the current $population$ and $sample$;}
\KwResult{$true$ is the $sample$ is representative of the $population$.}
\BlankLine
\BlankLine
$z$ $\leftarrow 100 * e^{-\lambda t}$\;
$n_\infty$ $\leftarrow \dfrac{z^2 p(1-p)}{e^2}$\;
\BlankLine
$n$ $\leftarrow \dfrac{n_\infty}{1 + \dfrac{n_\infty - 1}{|population|}}$\;
\BlankLine
\If{$|sample| > n$}{
    \If{t-test(population, sample, z)}{
        balanced $\leftarrow$ false\;
        \ForEach{request in population}{
            $E_{p}\leftarrow\{\, x \in population \mid x.id = request.id \,\}$\;
            $E_{s}\leftarrow\{\, y \in sample \mid y.id = request.id \,\}$\;
            \BlankLine
            \If{$\left(\dfrac{|E_{p}|}{|population|} = \dfrac{|E_{s}|}{|sample|}, z\right)$}{
                    \BlankLine
                    $balanced \leftarrow$ true\;
            }
        }
        \If{balanced}{
            release $sample$ for analysis\;
            $sample \leftarrow \emptyset$\;
            $population \leftarrow \emptyset$\;
            return $true$\;
        }
    }
}
return $false$\;
\caption{Sample Evaluation}
\label{algorithm:sampling-evaluation}
\end{algorithm}

% Ref: https://www.cuemath.com/exponential-decay-formula/
To prevent a scenario in which the sample being collected is never representative, Algorithm~\ref{algorithm:sampling-evaluation} uses an exponential \emph{decaying confidence} (line 1). It is used to adjust the required level of representativeness of the sample based on the length of the monitoring cycle. The confidence ($z$) starts at 100\% and decreases by a constant $\lambda$ every second ($t$) in the monitoring cycle, where $\lambda = 1 / maxLength$. \textit{maxLength} is the maximum length of the monitoring cycle, which is the required parameter of this activity.

The first criterion---sample size---is checked in line 4. The required sample size (line 3) is given by Cochran's minimum sample size~\cite{Cochran1977} (line 2), with finite population correction, where the decaying confidence level $z$ is used to estimate the associated standard normal distribution (e.g., 1.96 for $z = 0.95$), $e$ is the margin of error ($e = 0.05$) and 
% Ref: https://www.tarleton.edu/academicassessment/documents/samplesize.pdf (Degree of variability)
$p$ is the degree of variability of the population, indicating how heterogeneous the population is. We use $p = 0.5$ because it does not assume that the population is homogeneous and leads to higher samples, being thus a conservative choice.

Given that we track the performance of application requests, we test the equivalence between the sample and population---stored in Table~\ref{tab:performance_collection_example}---using this measurement, which is the second criterion. This is done by verifying if the frequency distribution of the sample significantly differs from the postulated population mean using a one-sample parametric t-test (two-sided) (line 5) to test the null hypothesis that the sample mean is equal to the population mean, with a decaying confidence level (computed in line 1). The test compares the average values of the two data sets and determines if they came from the same population.

Finally, the third criterion is evaluated in lines 6--14, which checks the balance in the request distribution, measuring the over- or under-representation of requests in the sample compared to the population. This is similar to the comparison shown in Table~\ref{tab:frequency_distribution}. However, in line 10, the comparison between each request type in the sample and in the population considers a margin of error of \textit{z} (decaying confidence level). This evaluation aims to ensure that the sample is balanced according to the population to give confidence that all requests, even those that rarely happen, are present.
% Example:
% error = popProportion - (popProportion * decayingConfidenceFactor)
% (popProportion + error >= samProportion >= popProportion - error)

When the sample meets all these three criteria, it is said representative and is released for analysis. After releasing the sample for analysis, we reset the sample and population (lines 16 and 17), and a new monitoring cycle starts.
% !TEX root = ../main.tex

\section{Evaluation}
\label{sec:evaluation}

Having described our monitoring process in detail, we now evaluate it using applications of a widely known and used benchmark.

\subsection{Evaluation Settings}

\subsubsection{Research Questions and Metrics}\label{subsec:study_settings}

Our monitoring process aims at collecting representative samples of execution traces and, at the same time, keeping the performance overhead at an acceptable level. Therefore, the goal of evaluation is to assess these two aspects, which are the focus of our two research questions, listed as follows. Metrics used to answer each question are also detailed.

\begin{description}
	\item[\textbf{RQ1}] What is the \emph{performance impact} of our monitoring process?
	\begin{description}
		\item[TR] Throughput (average number of requests / second)
		\item[SR] Average sampling rate / second
	\end{description}
	\item[\textbf{RQ2}] What is the \emph{representativeness} of the samples of executions traces collected with our monitoring process?
	\begin{description}
		\item[RMSE] Root-mean-square error of memory consumption
	\end{description}
\end{description}

Whenever we monitor a software application, there is performance overhead. Therefore, in RQ1, we quantitatively assess this overhead by measuring the application throughput (TR). In addition, given that the monitoring overhead is proportional to the sampling rate, to better understand the throughput, we also measure the sampling rate (SR).

A dynamic sampling rate may have a negative effect on the representativeness of collected traces, making them less useful for understanding the application behavior for a particular goal, such as debugging. For not biasing our evaluation towards our approach, in RQ2 we do not use any criteria on which our monitoring process relies to assess sample representativeness (e.g. response times). Instead, we use a software characteristic---memory consumed by methods---that can be understood through monitoring, is straightforward to be collected, and can have its correctness evaluated. The measurement of memory consumed by methods may not be trivially obtained some execution platforms. However, there is a reliable and commonly used way of obtaining this measurement, i.e.\ computing the difference between the available memory before and after the method execution. Such measurement is enough to compare two different executions of an application in terms of memory footprint (i.e.\ the amount of memory used by the application during its execution), considering that the workload variations and environment configuration are exactly the same for both executions. As we are pursuing a representative sample of execution traces, it is expected that the memory footprint resulting from the collected traces is representative of the population (i.e.\ all the methods processed by the application). Thus, based on a ground truth (explained as follows), we can assess the error (RMSE) of the sample with respect to the average memory usage of each method. If RMSE is low, it indicates that the collected traces are reliable for debugging memory consumption, for example.

\subsubsection{Compared Approaches}

Our adaptive monitoring process (\texttt{ADP}) is compared to the main alternative approach~\cite{Hauswirth2004,Bronink2016}, which is a sampling rate inversely proportional to the workload given by the throughput (\texttt{INV}), and a practically used approach, which is uniform sampling (\texttt{UNI}). The selected uniform sampling rate is 50\%, which is the same used as initial (and desired) sampling rate for \texttt{ADP} and \texttt{INV}.
As a reference, we also execute our evaluation with two additional configurations. The first is no monitoring (\texttt{NOM}), in which there is no monitoring overhead and it thus provides the maximum (best) possible value for throughput. The second is full monitoring (\texttt{FUM}), in which every application request is recorded with execution traces. \texttt{FUM} gives thus the minimum (worst) possible throughput value and also serves as ground truth for calculating RMSE because it contains the memory measurements of all executions of each request made during a workload simulation (i.e.\ the population).

\subsubsection{Target Applications}

We use the DaCapo~\cite{DaCapo:paper,DaCapo:TR} benchmark suite for our evaluation. DaCapo provides various Java applications to evaluate approaches that focus on the execution environment of applications. Consequently, it does not explicitly provide extension points for customizing its execution. Because we need to simulate a workload with variation (not simply firing requests one after another) as well as instrument the applications, we selected a subset of five applications to be instrumented, described as follows.

\begin{description}
	\item[\texttt{cassandra}] Executes queries to recover documents from the NoSQL database management system Cassandra.
	%executes a JDBCbench-like in-memory benchmark, executing a number of transactions against a model of a banking application, replacing the hsqldb benchmark
	\item[\texttt{h2}] Executes SQL transactions against a model of a banking application on top of the H2 database.
    %Uses lucene to do a text search of keywords over a corpus of data comprising the works of Shakespeare and the King James Bible
	\item[\texttt{lusearch}] Executes search queries against the document search engine Lucene.
	% runs the daytrader benchmark via a Jave Beans to a GERONIMO backend with an in-memory h2 as the underlying database
	%A web application and benchmark built around the paradigm of an online stock trading system
	\item[\texttt{tradebeans}] Runs HTTP requests via Java Beans against a web application that simulates a stock trading system.   
	%transforms XML documents into HTML
	\item[\texttt{xalan}] Calls multiple times an XSLT processor for transforming XML documents into HTML pages.
\end{description}

The rationale for selecting these particular applications is that they are all distributed applications designed to process multiple requests in parallel, e.g.\ web requests or database queries, and are based on domains in which monitoring is valuable and difficult to control in terms of overhead. 

\subsubsection{Procedure}

% INGRID: This might be not really needed, but left as a comment for the record: we selected the method signature for event identification

We instrumented all the target applications (which are open source) to enable monitoring using aspect orientation. Aspects intercept method calls, and then there are four Java implementations (\texttt{ADP}, \texttt{INV}, \texttt{UNI}, and \texttt{FUM}) to decide whether an application request should be recorded and, if so, collect execution traces. \texttt{NOM} corresponds to the original version of the applications.

As introduced, our monitoring process requires a set of parameters. They consist of a single configuration for all target applications and not values that must be tuned for specific applications. The Sampling Rate Adaptation activity is triggered every 1s and the collection of performance baselines last 3s. The maximum and minimum sampling rates are 50\% (as in \texttt{UNI}) and 1\%, respectively. Finally, the maximum length of a monitoring cycle is 180s (3 min).

For generating application workloads, we use the workload simulation provided by DaCapo, which relies on a navigation pattern that either falls into a specific distribution (transition table) or follows a specific sequence of executions. We use this navigation pattern to simulate a varying number of simultaneous users (i.e.\ threads). This allows us to observe the impact of monitoring when the application is under various stress conditions and how the sampling rate is adjusted. Inspired by load intensity modelling approaches~\cite{Kistowski2017}, the designed workload includes (a) situations in which it keeps a stationary number of users, (b) seasonal patterns, and (c) bursts in the number of simultaneous users. Therefore, the resulting workload covers the typical variations regarding usage scenarios that may occur in production environments. The same workload settings are used to execute each application with each compared approach.

The simulations were executed on an Intel i7 2GHz with 16G RAM. The maximum heap size of the Java Virtual Machine (JVM) was limited to 4GB to cause the applications to execute under stress (with limited resources considering the workload). Each simulation was executed 10 times. The maximum number of simultaneous users was selected based on the identification of which number of users causes the application to deteriorate its performance due to the lack of resources. With 4GB of RAM available, this number varies from 6 to 200. The measurement used to compare application executions and answer RQ2---memory consumed by methods---is not explicitly made available by the JVM. Thus, we use the discussed standard way to derive it, i.e.\ computing the difference between the available memory before and after the method execution. Given that there are moments in which JVM garbage collector is executed to free up memory, some of the collected measurements are negative and thus invalid. These are, therefore, discarded.

The error used in RMSE in our evaluation is given by the difference between the average memory consumption (in kilobytes) of all executions of a particular request (obtained with \texttt{FUM}) and the average memory consumption of executions monitored by a compared approach (\texttt{ADP}, \texttt{INV}, and \texttt{UNI}) of this request. RMSE is thus calculated as follows.
\begin{equation*}
    RMSE = \sqrt{\frac{\sum_{r in Req} (\mu_{FUN} - \mu_{S})^{2}}{|Req|}}
\end{equation*}
where $Req$ is the set of possible requests, $\mu_{FUN}$ is the average of memory consumption of all executions of request $r$ (measured by $\mu_{FUN}$), $\mu_{S}$ is the average of memory consumption of  executions of request $r$ measured by the sampling approach $S \in \{ADP, INV, UNI\}$, and $|Req|$ is the number of possible requests.

\subsection{Results}

The results obtained following the procedure described above are presented in Table~\ref{tab:results}. It shows the values obtained for each metric (TR, SR, and RMSE) with each compared approach (\texttt{ADP}, \texttt{INV}, \texttt{UNI}) and reference values (\texttt{NOM} and \texttt{FUM}) for each target application. Because we run the simulation 10 times for each configuration, we present the mean and standard deviation. As can be seen, the results are consistent across all applications, even though the applications vary in nature. A Friedman's test showed that there is significant difference among the compared approaches, both for TR ($\chi^2(2) = 84.28$, $p < 0.001$) and RMSE ($\chi^2(2) = 74.68$, $p < 0.001$). Post-hoc analysis with pairwise comparisons using Nemenyi-Wilcoxon-Wilcox all-pairs test for a two-way balanced complete block design revealed that this is due to the differences among all approaches in both cases.

% !TEX root = ../main.tex

\begin{table*}[t]
\centering
\caption{Simulation Results: Comparison of the values obtained for the metrics Throughput (TR), Sampling Rate (SR), and Root-mean Square Error (RMSE).\label{tab:results}}
\begin{tabular}{lllll}
\toprule
 & \textbf{Monitoring} & \textbf{Throughput} & \textbf{Sampling Rate} & \textbf{Root-mean Square Error} \\ \midrule 

\multicolumn{1}{l}{\multirow{5}{*}{\rotatebox[origin=c]{90}{\textbf{cassandra}}}} 
 & NOM & 23180.4$\pm$488.6 & 0\% & --- \\ 
 & FUM & 16769.5$\pm$468.7 (-27.6\%) & 100.0\% & --- \\ \cmidrule(l){2-5}
 & ADP & 20763.9$\pm$666.2 (-10.4\%) & 48.7\%$\pm$1.3 & 496.6$\pm$42.9 \\
 & INV & 21112.7$\pm$520.6 (-8.9\%) & 29.2\%$\pm$2.5 & 699.9$\pm$57.0 \\
 & UNI & 18900.2$\pm$661.7 (-18.4\%) & 50.0\% & 651.6$\pm$75.2 \\ \midrule
 
\multicolumn{1}{l}{\multirow{5}{*}{\rotatebox[origin=c]{90}{\textbf{h2}}}} 
 & NOM & 1829.0$\pm$20.1 & 0\% & --- \\ 
 & FUM & 1197.1$\pm$30.2 (-34.5\%) & 100.0\% & --- \\ \cmidrule(l){2-5}
 & ADP & 1587.1$\pm$19.8 (-13.2\%) & 44.0\%$\pm$3.4 & 628.0$\pm$108.6 \\
 & INV & 1633.2$\pm$24.4 (-10.7\%) & 32.7\%$\pm$1.5 & 1291.9$\pm$118.7 \\
 & UNI & 1517.0$\pm$19.1 (-17.0\%) & 50.0\% & 1196.1$\pm$102.8 \\ \midrule
 
\multicolumn{1}{l}{\multirow{5}{*}{\rotatebox[origin=c]{90}{\textbf{lusearch}}}} 
 & NOM & 74376.7$\pm$213.8 & 0\% & --- \\ 
 & FUM & 49397.7$\pm$363.9 (-33.5\%) & 100.0\% & --- \\ \cmidrule(l){2-5}
 & ADP & 66267.2$\pm$324.2 (-10.9\%) & 41.6\%$\pm$2.2 & 1394.2$\pm$349.0 \\
 & INV & 70951.0$\pm$216.0 (-4.6\%) & 28.0\%$\pm$1.6 & 3010.8$\pm$450.0 \\
 & UNI & 57142.0$\pm$433.3 (-23.1\%) & 50.0\% & 2086.0$\pm$251.1 \\ \midrule
 
\multicolumn{1}{l}{\multirow{5}{*}{\rotatebox[origin=c]{90}{\textbf{tradebeans}}}} 
 & NOM & 1832.2$\pm$20.2 & 0\% & --- \\ 
 & FUM & 1204.6$\pm$17.9 (-34.2\%) & 100.0\% & --- \\ \cmidrule(l){2-5}
 & ADP & 1571.8$\pm$14.0 (-14.2\%) & 48.4\%$\pm$0.9 & 721.2$\pm$40.3 \\
 & INV & 1619.6$\pm$15.1 (-11.6\%) & 32.7\%$\pm$1.1 & 797.7$\pm$55.5 \\
 & UNI & 1512.8$\pm$22.6 (-17.4\%) & 50.0\% & 831.0$\pm$36.7 \\ \midrule
 
\multicolumn{1}{l}{\multirow{5}{*}{\rotatebox[origin=c]{90}{\textbf{xalan}}}} 
 & NOM & 266.8$\pm$1.0 & 0\% & --- \\ 
 & FUM & 182.6$\pm$1.7 (-31.5\%) & 100.0\% & --- \\ \cmidrule(l){2-5}
 & ADP & 239.1$\pm$1.5 (-10.3\%) & 47.5\%$\pm$0.9 & 111.7$\pm$9.1 \\ 
 & INV & 241.1$\pm$1.7 (-9.6\%) & 24.0\%$\pm$0.6 & 187.4$\pm$13.2 \\
 & UNI & 230.5$\pm$1.3 (-13.5\%) & 50.0\% & 135.0$\pm$10.3 \\

\bottomrule
\end{tabular}
\end{table*}

With respect to performance overhead (RQ1), as expected, \texttt{UNI} achieves the worst results, causing the throughput to be 13.5--23.1\% lower than \texttt{NOM}. The impact is approximately 50\% of \texttt{FUM}, as it collects execution traces of roughly half of the requests. \texttt{INV} has the lowest performance overhead, with an overhead ranging from 4.6\% to 11.6\%. This occurs because it always reduces the sampling rate with more intense workloads, regardless of its impact on the collected execution traces. \texttt{ADP}, in turn, is the ``middle option'', which has an overhead from 10.3\% to 14.2\%, as it also takes sample representativeness into account while monitoring the application. 

The performance overhead is in accordance with the sampling rate. The lower the performance overhead, the lower the sampling rate. Despite achieving the intermediate results, in all cases, \texttt{ADP} has a performance overhead closer to \texttt{INV} than to \texttt{UNI}. Nevertheless, its average sampling rate is, also in all cases, closer to \texttt{UNI} than to \texttt{INV}, sometimes as high as 48.7\% (note that, by configuration, the sampling rate is always limited to 50\%). This indicates that \texttt{ADP} is able to choose the moments in which the sampling rate should be reduced (this is further discussed in the next section) as well as to reduce the sampling rate in a sustainable manner.

With respect to the error present in collected samples (RQ2), \texttt{ADP} is not the middle option. In all cases, it has the lowest (best) results for RMSE. This provides evidence that \texttt{ADP} is able to collect execution traces that better represent the population. Although, as expected, \texttt{INV} has the highest error for most applications, this is not the case for \texttt{tradebeans}. A possible explanation is that the memory consumption of the different application requests largely varies for this application and, in this particular case, relying on randomization to collect traces, even with higher sampling rates, cannot guarantee good results.

\begin{framed}
\noindent \textbf{Conclusion.} \texttt{ADP} is able to collect the most representative samples of execution traces, using memory consumption as representativeness measure. The error of the collected samples is 9--54\% and 12--44\% lower than \texttt{INV} and \texttt{UNI}, respectively. It also significantly reduces the performance overhead of \texttt{UNI} (3--12\% lower). Although it has a performance overhead higher than \texttt{INV}, it is much lower (1--6\%) than the reduction of the error in the collected samples.
\end{framed}

\subsection{Detailed Analysis}

\begin{figure*}[t]
    \centering
    \begin{subfigure}{0.49\linewidth}
        \includegraphics[width=\linewidth]{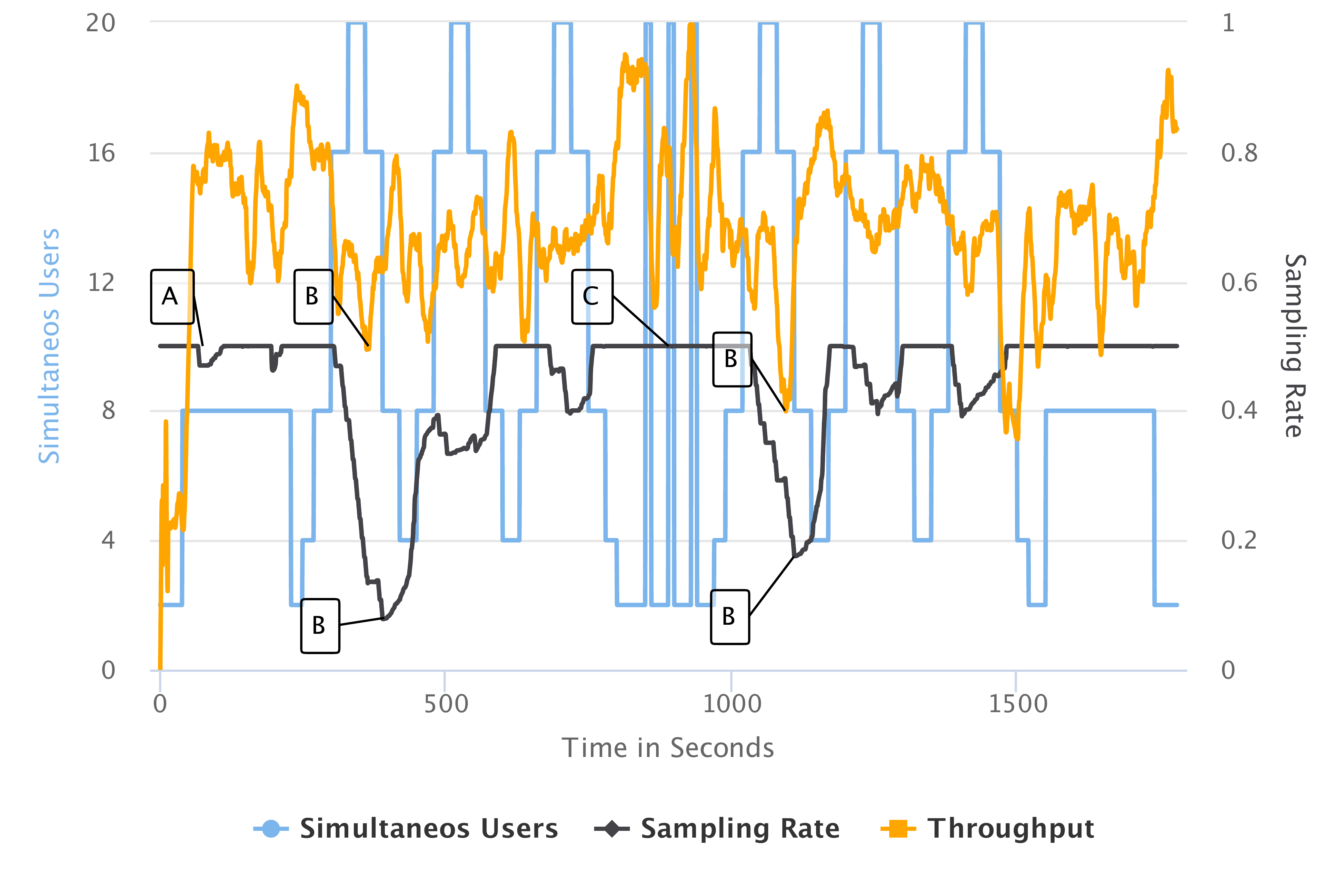}
        \caption{Interaction among workload, throughput, and sampling rate.\label{fig:SRvsTR}}
    \end{subfigure}
    \begin{subfigure}{0.49\linewidth}
        \includegraphics[width=\linewidth]{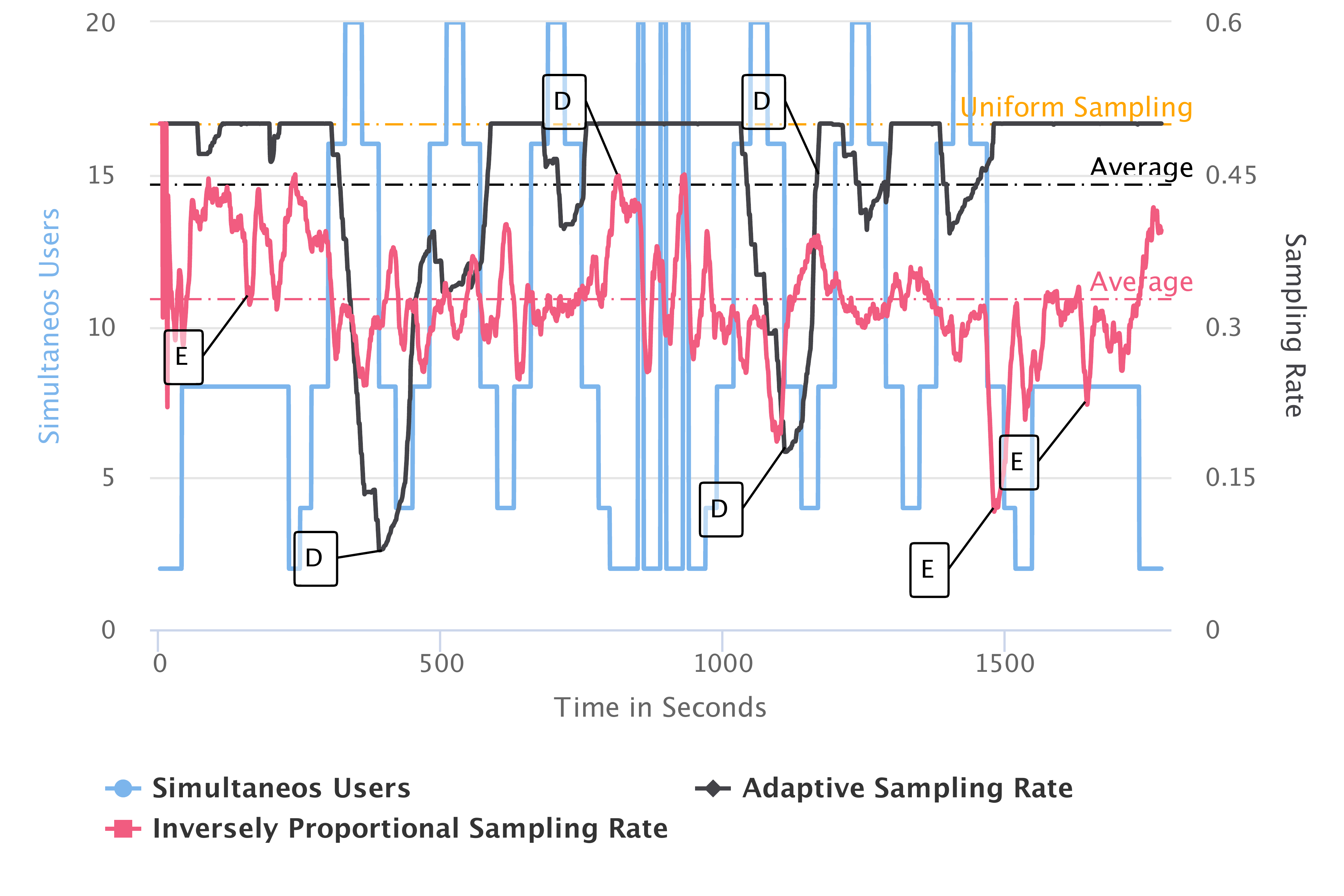}
        \caption{\texttt{ADP} Sampling Rate vs.\ \texttt{INV} Sampling Rate.\label{fig:ADPvsINV}}
    \end{subfigure}
    \begin{subfigure}{0.49\linewidth}
        \includegraphics[width=\linewidth]{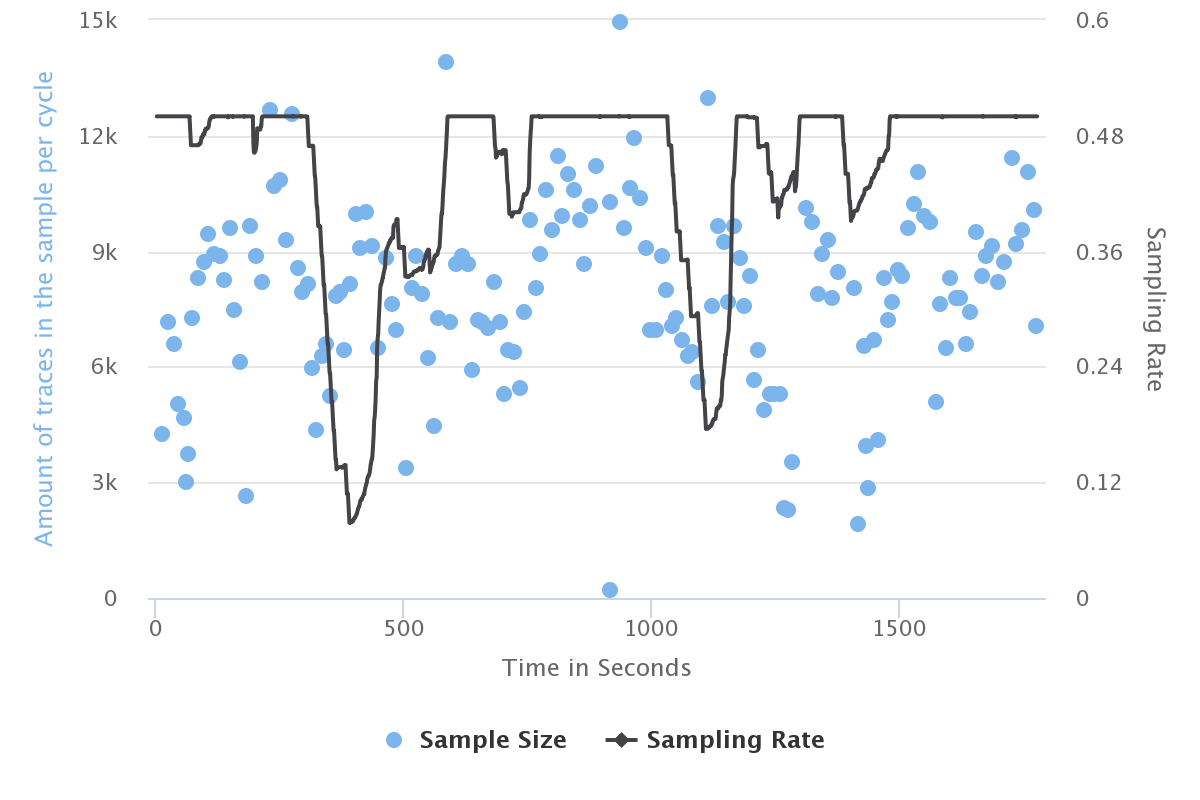}
        \caption{Collected sample sizes over time.\label{fig:sampleSize}}
    \end{subfigure}
    \begin{subfigure}{0.49\linewidth}
        \includegraphics[width=\linewidth]{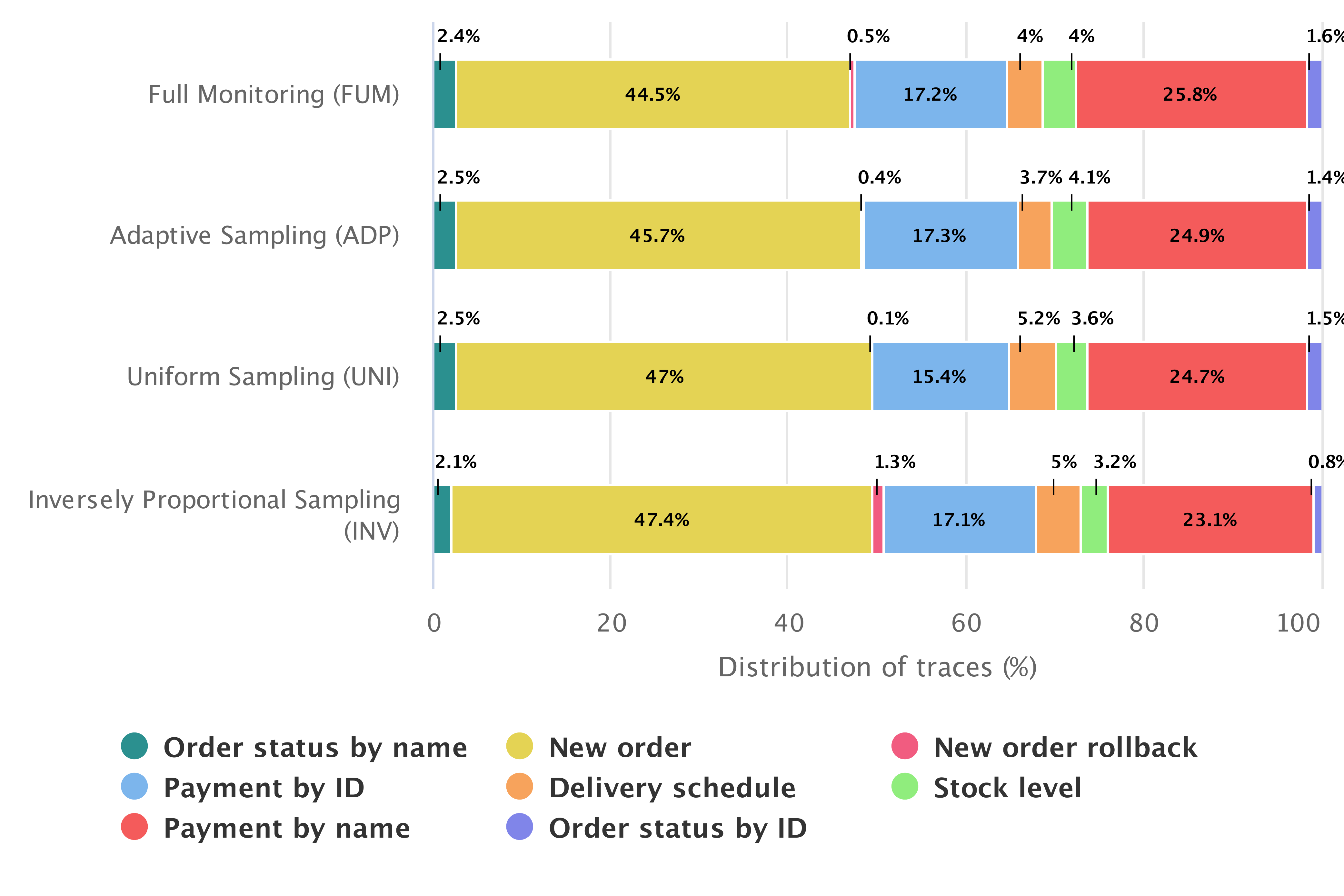}
        \caption{Comparison of the distribution of application requests.\label{fig:traceDistribution}}
    \end{subfigure}
    \caption{Analysis of the \texttt{ADP} Results with the \texttt{h2} Application.}
    \label{fig:h2Results}
\end{figure*}

In the previous section, the results show that \texttt{ADP} significantly improves the representativeness of the collected samples of execution traces, with little impact on the performance overhead. To explain these results, we analyze in detail the results obtained with \texttt{h2}, shown in Figure~\ref{fig:h2Results}.\footnote{All applications have similar results. Due to space restrictions, their charts are available in our complementary material at \url{https://www.inf.ufrgs.br/prosoft/resources/2022/tse-adaptive-sampling}.}
From the 10 executions, we selected that with the median throughput value.

\paragraph*{\textbf{Interaction among Workload, Throughput, and Sampling Rate}}

We first analyze what happens over the course of the simulation in Figure~\ref{fig:SRvsTR}. The blue line shows the application workload. The typical workload is 8 simultaneous users, which can be 20 in peeks (recall that the memory limit is 4GB, which causes the application to be under stress with a relatively low number of users). As explained, the workload has stationary segments, seasonal patterns and bursts. In stationary segments, \texttt{ADP} is able to keep the sampling rate at a value close to 50\% (label A), with small decreases due to variances in the response time, as this is the metric used to adapt the sampling rate. In seasonal patterns, \texttt{ADP} detects performance degradation and reduces the sampling rate (label B). Note that even with a decreased sampling rate, the throughput (orange line) decreases, showing that the user requests are causing the application to be under stress. Lastly, in isolated bursts, the sampling rate remains at 50\% (label C) because the increased number of users for brief moments does not have a major impact on the application performance. As can be seen, despite the monitoring and the peeks, the throughput is not lower than in the rest of the simulation.

\paragraph*{\textbf{\texttt{ADP} Sampling Rate vs.\ \texttt{INV} Sampling Rate}}

Now we look in detail at the sampling rate controlled by \texttt{INV} and how it differs from \texttt{ADP}. Both approaches apply mechanisms to reduce the sampling rate when the application is struggling with an increased workload (label D). However, while \texttt{ADP} uses the response time to make decisions, \texttt{INV} relies on the workload (throughput). In many cases, this correctly reduces the sampling rate to not cause a major performance impact on the application. But in certain situations (label E), low throughput is due to a low number of requests, thus there is no need to reduce the sampling rate. \texttt{ADP} is able to better understand the application as a whole as it keeps track of a performance baseline with and without monitoring, allowing it to identify when the monitoring is competing for resources with the application.

% PARA INGRID: cycles é média. 211 execution traces é events. 
\paragraph*{\textbf{Collected Sample Sizes}}

\texttt{ADP} does not focus on collecting execution traces to be analyzed all together, but works in cycles providing a set of samples of execution traces, each being representative of the population in each cycle. Although there is a timeout for cycles, ideally the cycle ends when the representativeness criteria are met, leading to samples of various sizes. We present the collected samples sizes for \texttt{h2} in Figure~\ref{fig:sampleSize}. The horizontal proximity between the dots indicates that no cycle reached the timeout of 180s---the maximum cycle time is 25 seconds. Figure~\ref{fig:sampleSize} also shows that the sample size is not correlated to the sampling rate. This may occur in \texttt{h2} due to the low number of types of requests (8 distinct types) because it is easier to have similar distributions when the number of classes to be compared between the sample and population is low. Note that there is an outlier cycle composed only of 211 execution traces and that lasted less than 1s. This indicates that the sample satisfied the representativeness criteria with high confidence because the longer the cycle, the lower the confidence level as it decays over time. As result, on average, \texttt{h2} had 163 cycles. Because the request types of \texttt{xalan} and \texttt{tradebeans} are also low, 16 and 12, respectively, it is possible to collect representative samples in shorter times, resulting in 422 and 151 cycles, respectively. \texttt{cassandra} and \texttt{lusearch}, in turn, have more than 100 request types, causing the lowest number of cycles (122 and 48, respectively). We observed longer monitoring cycles in these two applications, including timeouts.

\paragraph*{\textbf{Distribution of Application Requests}} 

Lastly, we analyze the distribution of application requests in Figure~\ref{fig:traceDistribution} considering the data in all samples collected during one execution of the simulation of \texttt{h2}. \texttt{FUM} shows the distribution of the population (ground truth). Although \texttt{ADP} checks for distribution similarity by sample, the resulting set of samples has the distribution most similar to the population (considering the whole simulation), having the \texttt{new order} request the highest difference (1.2\%). This request type, which is the most frequent, also led to the highest difference for \texttt{UNI} and \texttt{INV}. \texttt{UNI} has a difference of 2.5\%; while \texttt{INV}, which focuses on performance rather than representativeness, has the highest difference (2.9\%) among the three approaches.

\subsection{Threats to Validity} 

Our evaluation involves runtime execution with a particular workload and, thus, there are many settings that may influence the results. All our settings were selected to avoid bias. The fired application requests have a key role in the obtained results. To minimize the chance of using a workload that favors a particular approach, we rely on the randomness and reliability provided by DaCapo. Another workload configuration that may influence the results is the number of simultaneous users and how it varies over time. Our designed workload includes different types of variations, which are those used in previous work. Moreover, the maximum number of users, based on preliminary executions, was selected to guarantee that the application executes under stress in certain moments. Another construction threat to validity is how we assess representativeness. The key goal is to evaluate whether the desired execution traces are included in the sample. Given that this depends on the monitoring goal, we use memory usage due to the reasons explained in the study settings. This measurement is not used by any of the compared approaches for adapting the sampling rate or making decisions. The only challenge is to collect this information in Java, because its virtual machine offers limited support to fine-grained memory measurements and, in addition, it has multiple features that can affect this kind of measurement during the application execution, such as the garbage collector and just-in-time compilation. We used a standard way to measure memory usage as well as discarded invalid measurements---negative values due to the execution of the garbage collector---for all approaches, including \texttt{FUM}.

Finally, an external threat to validity is the set of target applications. We selected applications that use various technologies, are of different domains, and vary in processing nature, e.g.\ while some make extensive memory usage, others rely on I/O or processor. Though the number of applications is not large, we emphasize that the obtained results are consistent across all applications and thus provide evidence of the generalization of the results. Yet, as any empirical study, further evaluations with different settings would improve the generality and reliability of the results.

\subsection{Limitations} 

We now point out limitations of our monitoring process.
% %limitação: the sample may take too much time to be collected, it may reduce the chance to find adaptation opportunities
% %Ameaça/limitação: sample sempre corresponde com o ciclo anterior - mas a adaptação cobre isso parcialmente
A monitoring cycle finishes when the representativeness criteria are met. In situations that an application must timely react to particular requests, this may cause the application to give a delayed response. We addressed this issue using a decaying confidence level based on the monitoring cycle time frame, which can be customized. However, the more elapsed time, the lower the confidence level. Therefore, if an application requires samples with some confidence level guarantees, the sample evaluation activity must be adapted.

% %limitação: coarse-grained still imply overhead - we can use sampling on it too. We could also use bootstrapping to generate population from the samples we have.
In our work, we monitor applications by continuously making decisions and adaptations to collect execution traces, which implies an overhead higher than making simple adjustments (as in \texttt{INV}). This is, however, done in a lightweight way and our evaluation showed that despite the execution of our process activities, we obtain the most representative samples with a performance not far from \texttt{INV}. Yet, it is possible to reduce the cost of the Sampling Rate Adaptation activity by using bootstrapping and other statistical techniques to generate data from samples and estimate the population based on monitoring time frames instead of instrumenting all the requests. Then, the monitoring can be disabled for extended periods when the sample is in good shape to be used in bootstrapping.

% !TEX root = ../main.tex

\section{Conclusion}
\label{sec:conclusion}

Software runtime monitoring has been largely used for a wide range of purposes, from debugging to self-adaptation. When it collects costly information like detailed execution traces in production environments, it is crucial to prevent the monitoring to cause unacceptable overhead. A typical approach is thus to sample the traces. However, it is important to pursue that the collected traces are representative of the population of execution traces.

In this paper, we proposed a monitoring process to find the sweet spot between these conflicting goals, i.e.\ overhead vs.\ representativeness. Our process is performed in monitoring cycles and is composed of three activities, which use algorithms with statistical foundations to decide whether a particular application request must be recorded, when and to what degree adapt the sampling rate, and determine when a sample has been collected to, then, begin a new monitoring cycle. We evaluated our process by comparing it with monitoring performed with uniform sampling and a sampling rate that is inversely proportional to the workload (\texttt{INV}) as well as used executions with no monitoring and monitoring every application request as a reference. Our results show that our approach collects samples with the lowest errors with respect to the population, having a performance impact that is only 1--6\% higher than \texttt{INV}, which achieves the highest error rate. 

Our approach is language independent. However, it was implemented in Java for our empirical evaluation. Our future work involves implementing and evaluating our approach in projects written in other programming languages. In addition, as monitoring can be used for different purposes, we aim to assess the effectiveness of our monitoring process for post-mortem fault analysis. Finally, we aim to adapt the process for distributed architectures.

\section*{Acknowledgments}
Ingrid Nunes thanks for CNPq grants ref. 313357/2018-8 and ref. 428157/2018-1. This study was financed in part by the Coordena\c{c}\~{a}o de Aperfei\c{c}oamento de Pessoal de N\'{i}vel Superior - Brasil (CAPES) - Finance Code 001.

\bibliographystyle{elsarticle-num-names} 
\bibliography{references}
\end{document}